# Visualizing Multiphase Flow And Trapped Fluid Configurations In A Model Three-Dimensional Porous Medium

Amber T. Krummel[1,3,†], Sujit S. Datta[1,†], Stefan Münster[1,2], and David A. Weitz[1,*]

† These authors contributed equally to this work.

1 – Department of Physics, Harvard University, Cambridge MA 02138

2 – Max Planck Institute for the Science of Light and Center for Medical Physics and Technology, Universitat Erlangen-Nürnberg, Erlangen, Germany

3 – Current address: Department of Chemistry, Colorado State University, Fort Collins CO 80523

* Email: weitz@seas.harvard.edu

## Abstract

We report an approach to fully visualize the flow of two immiscible fluids through a model three-dimensional (3D) porous medium at pore-scale resolution. Using confocal microscopy, we directly image the drainage of the medium by the non-wetting oil and subsequent imbibition by the wetting fluid. During imbibition, the wetting fluid pinches off threads of oil in the narrow crevices of the medium, forming disconnected oil ganglia. Some of these ganglia remain trapped within the medium. By resolving the full 3D structure of the trapped ganglia, we show that the typical ganglion size, and the total amount of residual oil, decreases as the capillary number $Ca$ increases; this behavior reflects the competition between the viscous pressure in the wetting fluid and the capillary pressure required to force oil through the pores of the medium. This work thus shows how pore-scale fluid dynamics influence the trapped fluid configurations in multiphase flow through 3D porous media.

## Topical Heading and Key Words

Fluid mechanics and transport phenomena; porous media, multiphase flow, permeability, capillarity, wetting.

**Introduction**

Multiphase flow through porous media is important for a diverse range of applications, including aquifer remediation, $CO_2$ sequestration, and oil recovery[1]. These often involve the displacement of an immiscible non-wetting fluid from a porous medium by a wetting fluid, a process known as imbibition. The random structure of the pore space typically leads to complex fluid displacement through the pores[2-5]; consequently, imbibition can lead to the formation of disconnected ganglia of the non-wetting fluid[6]. Some ganglia can be mobilized and removed from the medium; however, many ganglia become trapped within it. This phenomenon may be particularly important in oil recovery, where over 90% of the oil within a reservoir can remain trapped after primary recovery. The complex flow behavior leading to non-wetting fluid displacement and trapping has been visualized in two-dimensional (2D) micromodels[7,8]; however, a complete understanding of the physics underlying the formation and trapping of ganglia requires experimental measurements on three-dimensional (3D) porous media[9-12].

Optical techniques typically cannot be used to directly image the flow through such media due to the light scattering caused by the differences in the indices of refraction when multiple fluid phases are used. Instead, magnetic resonance imaging (MRI)[13,14] and X-ray micro computer tomography (X-ray $\mu$CT)[15-18] have been used to visualize either the bulk flow dynamics, or the individual ganglia, within 3D porous media; however, fast visualization at pore-scale resolution is typically challenging. As a result, despite its broad technological importance, the dependence of ganglion formation and trapping on the pore-scale flow conditions remains unclear. Our understanding of the physics underlying these processes can be improved by a combination of direct visualization of the multiphase flow at pore-scale resolution with 3D characterization of the resulting trapped ganglia configurations.

In this article, we report an approach to visualize the pore-scale dynamics of ganglion formation and trapping, and to characterize the intricate structure of the trapped ganglia, within a 3D water-wet porous medium. We match the refractive indices of the wetting fluid, the non-wetting oil, and the porous medium; this enables us to directly image the structure of the medium, and the multiphase flow within it, in 3D using confocal microscopy. We find that the wetting fluid displaces oil from the medium by flowing along the solid surfaces, pinching off threads of oil in the narrow crevices between them, before displacing oil from the pores; consequently, the flow is highly non-local and some oil ganglia remain trapped within the pores. During oil displacement, the oil pinch-off is diminished for increasing capillary number *Ca*. Our experimental approach enables us to fully resolve the 3D structure of the resulting trapped ganglia. We show that the ganglia configurations are strongly dependent on flow history: the typical ganglion size and the total amount of residual oil decrease as $Ca$ increases. Our results indicate that the geometry of the trapped oil is determined by the competition between the viscous pressure in the wetting fluid during oil displacement and the capillary pressure required to force oil through the pores of the medium. This work thus shows how pore-scale fluid dynamics influence the trapped fluid configurations in multiphase flow through porous media.

**Experimental Methodology**

We prepare rigid 3D porous media by lightly sintering[19] densely-packed hydrophilic glass beads, with polydispersity ≈ 4%, in thin-walled rectangular quartz capillaries [Figure 1(a)]; these have cross-sectional areas $A \approx$ 1mm x 1mm or 1mm x 3mm. The beads have average radius $a$ = 75 $\mu$m or 32 $\mu$m; the media thus have lateral dimensions spanning from approximately 7 to 50 beads[*]. Scattering of light from the interfaces between the wetting and non-wetting fluids, as well as from the interfaces between the fluids and the beads, typically precludes direct observation of the multiphase flow in 3D. We overcome these limitations by matching the refractive indices of the wetting fluid, the non-wetting oil, and the beads, enabling full visualization of the multiphase flow in 3D[20-25,†]. We formulate a wetting fluid comprised of a mixture of dimethyl sulfoxide and water at 91.4% and 8.6% by weight, respectively; to visualize this fluid using confocal microscopy, we add 0.01 vol% fluorescein dye buffered at pH = 7.2. Additionally, we formulate another non-wetting fluid comprised of a mixture of aromatic and aliphatic hydrocarbon oils (Cargille Labs). These mixtures are designed to closely match the refractive indices of the wetting fluid and the non-wetting fluid to each other and to the refractive index of the glass beads. The wetting and non-wetting fluids have densities $\rho_w$ = 1.1 g/cm$^3$ and $\rho_{nw}$ = 0.83 g/ cm$^3$, respectively; we note that the flow through the porous medium is horizontal. The interfacial tension between the fluids is $\gamma$ = 13.0 mN/m, as measured using a du Noüy ring; this value is similar to the interfacial tension between crude oil and water[26]. The viscosities of the wetting and non-wetting fluids are $\mu_w$ = 2.7 mPa-s and $\mu_{nw}$ = 16.8 mPa-s, respectively, as measured using a strain-controlled rheometer; thus, our experiments are characterized by a viscosity ratio $M \equiv \mu_w/\mu_{nw} \approx 0.2$. We use confocal microscopy to estimate the three-phase contact angle made between the wetting fluid and a clean glass slide in the presence of the non-wetting fluid, $\theta \approx 5°$. For some of the results reported here, we use a wetting fluid comprised of a mixture of dimethyl sulfoxide, benzyl alcohol, ethanol, and water at 55.8%, 21.4%, 10.4%, and 12.3% by weight, respectively, and a non-wetting fluid comprised of a different mixture of the aromatic and aliphatic hydrocarbon oils (Cargille); these proportions closely match the refractive index of the glass beads and the non-wetting fluid, and are characterized by $\rho_w = 1.05$ g/cm$^3$, $\rho_{nw} = 0.82$ g/cm$^3$, $\mu_w = 2.2$ mPa-s, $\mu_{nw} = 7.4$ mPa-s, and $\gamma = 27.3$ mN/m.

---

[*] Because of the limited lateral size of the porous media, we note that boundary effects may influence the results. Future work is required to elucidate the role played by boundaries. However, we observe similar behavior to that reported here for pores near the boundaries.

[†] Unlike other imaging approaches like MRI or X-ray μCT, our approach does not enable full 3D imaging of optically opaque systems in which the refractive indices of the fluids are not matched to that of the glass beads.

We instrument the porous media to enable measurement of the bulk transport properties simultaneously with flow visualization. We use a differential pressure sensor to measure the pressure drop $\Delta P$ across a porous medium prior to and during visualization of the flow using confocal microscopy. We vary the volumetric flow rate $Q$, and measure the proportionate variation in $\Delta P$; this enables us to determine the absolute permeability of a medium with length $L$ and cross-sectional area $A$, $k \equiv \mu_w(QL/A)/\Delta P$. The permeability of a disordered packing of spheres can be estimated using the Kozeny-Carman relation, $k = a^2\varphi^3/45(1-\varphi)^2$, where $\varphi$ is the porosity of the packing and $a$ is the average sphere radius. The dependence of the measured permeability on bead size is consistent with $k \sim a^2$, in agreement with this prediction. Moreover, for a medium with bead radius $a = 75$ $\mu$m and $\varphi = 41\%$, we expect $k = 25$ $\mu$m$^2$; we find $k \approx$ 75-95 $\mu$m$^2$, in reasonable agreement with our expectation. The discrepancy between the measured permeability and the theoretical prediction is unclear; it may, for example, arise from the effect of the capillary walls confining our porous media[§].

We exploit the close match between the refractive indices of the fluorescently-dyed wetting fluid and the glass beads to visualize the structure of the porous medium in 3D. Prior to each experiment, the porous medium is evacuated under vacuum and saturated with $CO_2$ gas; this gas is soluble in the wetting fluid, preventing the formation of any trapped gas bubbles. We then fill the medium with the fluorescently-dyed wetting fluid by imbibition; a similar approach is used to saturate a rock core prior to core-flood experiments. We use a confocal microscope to image a 2$\mu$m-thick optical slice spanning a lateral area of 911.8$\mu$m x 911.8$\mu$m within the medium, and identify the glass beads by their contrast with the wetting fluid, as exemplified by the slices shown in Figure 1(b). To visualize the pore structure in 3D, we acquire a 3D image stack of 122 slices, each spaced by 2$\mu$m along the $z$-direction, within the porous medium, as shown in Figure 1(c). We use these slices to reconstruct the 3D structure of the medium [Figure 1(c)]. The packing of the beads is disordered; to quantify the porosity, $\varphi$, of this packing, we integrate the fluorescence intensity over all slices making up a stack. To probe the spatial dependence of $\varphi$, we also image stacks at multiple locations along the length of the medium. We find $\varphi = 41 \pm 3\%$ independent of position along the length of the medium, as shown in Figure 2; this is comparable to the porosity of highly porous sandstone[27]. Moreover, $\varphi$ is similar for different realizations of a porous medium, as shown by the different symbols in Figure 2; this illustrates the reproducibility of our protocol.

---

[§] Consistent with this hypothesis, we find that the discrepancy between the measured permeability and the theoretical prediction is smaller for porous media with larger cross-sectional areas and comprised of smaller beads.

To mimic discontinuous core-flood experiments on reservoir rocks, we subsequently flow > 30 pore volumes of non-wetting oil at a prescribed volumetric flow rate $Q$ through the porous medium; this process is often referred to as primary drainage[2,9]. We then flow dyed wetting fluid at the same flow rate; this process is referred to as secondary imbibition. Our experiments are performed at controlled flow rates to investigate the influence of the capillary number $Ca \equiv \mu Q/A\gamma$ on the flow; this represents the ratio of viscous to capillary forces, with $\mu = \mu_{nw}$ during drainage and $\mu = \mu_w$ otherwise. Our experiments span the range $Ca \sim 10^{-6}\text{-}10^{-3}$.

**Results and Discussion**

To investigate the pore-scale dynamics of primary drainage, we use confocal microscopy to visualize oil invasion at a single 11$\mu$m-thick optical slice within a porous medium of width 3mm and height 1mm, acquiring a new image every 35 ms. Because the oil is undyed, we identify it by its contrast with the dyed wetting fluid in the measured pore space. At low $Ca \sim 10^{-6}\text{-}10^{-4}$, the oil menisci displace the wetting fluid through a series of abrupt bursts into the pores [Figure 3(a)]; this indicates that a threshold pressure must build up in the oil at a pore entrance before it can invade the pore[28-30,67]. This pressure is given by the pore-scale capillary pressure, $2\gamma/a_t \sim 10^4$ Pa, where $a_t \approx 0.18a$ is the radius of a pore throat[31-33]. The bursts are typically only one pore wide[34], but can span many pores in length along the direction of the local flow [third frame of Figure 3(a)]; moreover, the oil remains continually connected during flow. We find that bursts can proceed along directions other than the bulk flow direction; as a result, the interface between the invading oil and the wetting fluid is ramified. After oil invasion, we observe a ~1$\mu$m-thick layer of the wetting fluid coating the bead surfaces[2-5], as indicated in the rightmost panels of Figure 3; because we use optical imaging, we can resolve this layer to within hundreds of nanometers.

The speed of the fluid meniscus during a burst of oil into a pore, $v$, can be estimated by balancing the threshold capillary pressure, $2\gamma/a_t$, with the viscous pressure required to displace the wetting fluid over a length $l$, $\mu_w \varphi v l/k$. Our experimental approach enables us to directly visualize, and quantify the speed of, individual bursts. For a porous medium with $a = 75\mu m$, $k = 75\mu m^2$, $\varphi = 0.41$, and cross-sectional width $w = 3mm$, we measure a maximum burst speed $v \approx 10mm/s$; this corresponds to wetting fluid flow over $l \sim 10mm$. While the details of this flow are complex, this simple scaling estimate suggests that the wetting fluid is displaced over a length scale comparable to the width of the porous medium, spanning many pores in size[†]; interestingly, this observation is consistent with previous measurements of pressure fluctuations during drainage[35], as well as imaging of drainage through a monolayer of glass beads[67,‡].

To explore the dependence of the water displacement on flow conditions, we visualize primary drainage for varying $Ca$. Unlike the low $Ca$ case, the oil bursts are not successive during primary drainage at higher $Ca \sim 10^{-4}$-$10^{-2}$; instead, neighboring bursts occur simultaneously, typically in the bulk flow direction [Figure 3(b)]. As a result, over the scale of multiple pores, the interface between the invading oil and the wetting fluid is more compact; this behavior reflects the increasing contribution of the viscous pressure in the invading oil at higher $Ca$[28,37,67]. As in the low $Ca$ case, we do not observe evidence for oil pinch off or subsequent reconnection; interestingly, this behavior is in contrast to the prediction that the oil can be pinched off during drainage[38-40]. Similar to the low $Ca$ case, we observe a $\sim 1\mu m$-thick layer of the wetting fluid coating the bead surfaces after oil invasion[2-5], as indicated in the rightmost panels of Figure 3.

To mimic discontinuous core-flood experiments on reservoir rocks, we flow dyed wetting fluid immediately after oil invasion. During secondary imbibition at low $Ca \sim 10^{-6}$-$10^{-5}$, the wetting fluid does not flow directly into the pores. Instead, it pinches off threads of oil at multiple nonadjacent pore constrictions; the resultant state is shown in the first frame of Figure 4(a). This is in stark contrast to the case of drainage. Such behavior must require the wetting fluid to initially flow through the thin wetting layers coating the bead surfaces. This observation directly confirms predictions for water-wet 3D porous media[41-47]. As flow proceeds, the wetting fluid spreads from the filled constrictions, displacing oil from the surrounding pores[5,7,45,48-51] [second and third frames of Figure 4(a)]. Consequently, the flow is highly non-local. The wetting fluid eventually forms a tortuous, continuous network of filled pores through which it continues to flow, forming disconnected oil ganglia in the process. For sufficiently long times, we do not observe any additional oil displacement, and a significant amount of oil remains trapped in the medium [last frame in Figure 4(a)]. The pressure drop across the porous medium does not appreciably change, further confirming that a steady state is reached.

---

[†] The estimated $l$ is larger than the width of the medium, suggesting that that the wetting fluid may not only flow across the width of the medium, but along it, as well.

[‡] Unlike the experiments reported in Ref. 67, the Reynolds number associated with an individual burst is $Re = \rho_{nw} v a/\mu_{nw} \sim 10^{-2}$-$10^{-1}$, suggesting that viscous forces dominate inertial forces in our experiments.

To explore the dependence of oil displacement on flow conditions, we visualize secondary imbibition for varying $Ca$. Unlike the low $Ca$ case, we do not observe oil pinch-off at higher $Ca \sim 10^{-4}$-$10^{-3}$; instead, the wetting fluid displaces the oil from the pores, as shown in Figure 4(b). This indicates that flow through the thin wetting layers becomes less significant as $Ca$ is increased. This observation confirms the predictions of recent simulations[37,44]. However, some of the oil is still bypassed by the wetting fluid, forming disconnected oil ganglia[52]; in several cases, the ganglia break up into smaller ganglia. Many of these ganglia are mobilized from the medium; however, a few smaller ganglia remain trapped [last frame in Figure 4(b)]. For sufficiently long times, these ganglia cease to move, and the pressure drop across the medium does not appreciably change, indicating that a steady state is reached. Our results highlight the important role played by the wetting layers in influencing the flow behavior.

To understand why some ganglia remain trapped, we analyze the distribution of pressures in the wetting fluid as secondary imbibition proceeds, before the oil is mobilized from the medium. Because the oil occludes some of the pore volume, the permeability of the medium to the wetting fluid is modified by a factor $\kappa \sim 0.1$; we note that $\kappa$ increases as the oil saturation decreases, as confirmed by our independent measurements. We thus estimate the viscous pressure gradient in the wetting fluid as $\mu_w(Q/A)/\kappa k$; for displacement at low $Ca$, this gradient is ~10 Pa/pore. Thus, for low $Ca$, the viscous pressure during oil displacement becomes comparable to the capillary pressure required to force oil through the medium, $2\gamma/a_t$ ~ $10^4$ Pa, only on length scales larger than ~$10^3$ pores. As a result, oil ganglia smaller than ~$10^3$ pores cannot be mobilized; we thus expect many large ganglia to remain trapped in the medium, consistent with our observations [last frame in Figure 4(a)]. By contrast, the viscous pressure gradient in the wetting fluid can be as large as ~$10^3$ Pa/pore for the highest $Ca$ studied. Consequently, the viscous pressure during oil displacement becomes comparable to the capillary pressure on length scales larger than ~10 pores. As a result, all ganglia larger than ~10 pores can be mobilized; we thus expect only a few smaller ganglia to remain trapped in the medium, consistent with our observations [last frame in Figure 4(b)].

To further test this picture, we exploit the close match between the refractive indices of the fluorescently-dyed wetting fluid, the undyed non-wetting fluid, and the glass beads to directly visualize the pore-scale configurations of the trapped oil. We image a second 3D image stack of 2-$\mu$m thick slices, and identify the undyed oil by its additional contrast with the dyed wetting fluid in the measured pore volume. By comparing the optical slices to slices taken at the same positions within the medium prior to primary drainage, we resolve the full 3D structure of the trapped oil ganglia; two representative examples are shown in Figure 5. The spatial resolution of this approach is on the order of hundreds of nanometers, significantly better than the typical limits of MRI and X-ray $\mu$CT. Interestingly, the ganglia sizes and shapes are highly dependent on $Ca$. At low $Ca \sim 10^{-6}$-$10^{-5}$, the trapped ganglia are ramified and can span many pores, as shown in Figure 5(a). By contrast, ganglia produced at higher $Ca \sim 10^{-4}$-$10^{-3}$ are typically smaller and less ramified, as shown in Figure 5(b).

To explore the variation of ganglia configurations with flow conditions, we use the 3D reconstructions to measure the volume of each oil ganglion visualized. Moreover, we image additional stacks at multiple locations along the length of the medium, resolving a total of over 500 individual ganglia for each $Ca$ investigated. We summarize these data by calculating the cumulative probability distribution function of the ganglion volume for each $Ca$[53,54]. At the lowest $Ca \sim 10^{-6}$-$10^{-5}$, the ganglia typically occupy tens of pores. The largest ganglia occupy at most several hundred pores [rightmost curves in Figure 6(a)], consistent with our expectation that ganglia larger than $\sim 10^3$ pores are mobilized from the medium. As $Ca$ increases, the viscous pressure in the wetting fluid during oil displacement increases; consequently, even smaller ganglia are mobilized and removed from the medium. Consistent with this expectation, we find that the median ganglion volume decreases with increasing $Ca$, in agreement with previous work[51,55-58]. Indeed, in stark contrast to the low $Ca$ case, ganglia formed at the highest $Ca \sim 10^{-4}$-$10^{-3}$ typically occupy only a few pores. The largest ganglia occupy at most 10 pores [leftmost curves in Figure 6(a)], in good agreement with our expectation that ganglia larger than $\sim 10$ pores are mobilized from the medium. These results support the idea that oil mobilization and trapping is determined by the interplay between the macroscopic viscous and capillary pressures[67].

To further quantify the properties of the trapped oil, we use the 3D reconstructions to calculate the residual oil saturation $S_{or} = V_{oil}/\varphi V$, where $V_{oil}$ is the total volume of oil trapped within the entire volume of porous medium imaged, $V$, for each $Ca$ investigated. We find that $S_{or} \approx 15\%$ for the lowest $Ca \sim 10^{-6}$-$10^{-5}$, in agreement with previous measurements[59]. Interestingly, $S_{or}$ decreases precipitously as $Ca$ increases above a threshold ~2 x $10^{-4}$, reaching ≈ 5% at the highest $Ca$ ~ $10^{-3}$ [Figure 6(b)]. This behavior reflects the combined effect of the diminished ganglion formation, and the enhanced mobilization of oil by the wetting fluid[59,60,68], as $Ca$ increases. In particular, as $Ca$ increases above this threshold, the viscous pressure in the wetting fluid balances the capillary pressure required to mobilize oil on length scales approaching the scale of an individual pore, consistent with theoretical predictions[68].

**Conclusion**

The experimental approach reported here provides a way to investigate both the pore-scale dynamics of ganglion formation and trapping, and the complex configurations of the trapped ganglia, within a 3D porous medium. By matching the refractive indices of the wetting fluid, the non-wetting oil, and the porous medium, we fully visualize the multiphase flow in 3D. We use confocal microscopy to directly visualize the drainage of the medium by a non-wetting oil and subsequent imbibition by a wetting fluid. During imbibition, we find that the wetting fluid can flow through thin layers coating the solid surfaces of the medium, pinching off threads of oil in the narrow crevices. This non-local flow forms disconnected oil ganglia, some of which remain trapped within the medium. During oil displacement, the oil pinch-off is diminished for increasing capillary number $Ca$. Moreover, the viscous pressure due to wetting fluid flow can mobilize increasing amounts of oil at higher $Ca$. Consequently, both the typical ganglion size, and the total amount of residual oil, decrease as $Ca$ increases. Our observations thus highlight the critical role played by pore-scale fluid dynamics in determining the trapping and mobilizing of oil in 3D porous media. These results may also be relevant to many other technologically-important multiphase flows, such as groundwater contamination by non-aqueous pollutants[61,62] and the storage of super-critical $CO_2$ within saline aquifers[63-66].

**Acknowledgement**

It is a pleasure to acknowledge D. L. Johnson, S. H. Kim, H. A. Stone, and the AEC consortium for useful discussions, and the anonymous reviewers for valuable feedback on the manuscript. This work was supported by the AEC and the Harvard MRSEC (DMR-0820484). SSD acknowledges support from ConocoPhillips.

**Figure Captions**

**Figure 1. Overview of the experimental approach.** (a) Schematic illustrating the structure of a typical 3D porous medium and imaging of flow within it using confocal microscopy. (b) Three optical slices, of thickness 2.0$\mu$m and lateral area 911.8$\mu$m x 911.8$\mu$m, taken at three different depths within a medium comprised of beads with average radius $a = 75\mu$m. The medium has been saturated with dyed wetting fluid; the black circles show the beads, while the bright space in between shows the imaged pore volume. (c) 3D reconstruction of a porous medium comprised of beads with average radius $a = 75\mu$m, with cross-sectional area 910$\mu$m x 910$\mu$m; the image shows the reconstruction of a section 350$\mu$m high for clarity.

**Figure 2. Porosity $\varphi$ of porous media is the same for different positions and realizations.** We find $\varphi = 41 \pm 3\%$ at multiple positions along the length of the porous media, and for different porous media prepared in the same way (different symbols).

**Figure 3. Pore-scale dynamics of primary drainage depend strongly on *Ca*.** Images show multiple frames, taken at different times, of a single optical slice. The slice is 11$\mu$m thick and is imaged within a porous medium comprised of beads with average radius $a = 75\mu$m, with cross-sectional width 3mm and height 3mm. The first frame in each sequence shows the imaged pore space, saturated with dyed wetting fluid; the dark circles are the beads, while the additional dark areas show the invading undyed oil. The beads, and the saturated pore space in between them, are subtracted from the subsequent frames in each sequence; thus, the dark areas in the subsequent frames only show the invading oil. Direction of bulk oil flow is from left to right. The final frame shows the unchanging steady state. Labels show time elapsed after first frame. (a) At low $Ca = 6 \times 10^{-5}$, the oil menisci displace the wetting fluid through a series of abrupt bursts into the pores, and the invading fluid interface is ramified over the scale of multiple pores. (b) At high $Ca = 4 \times 10^{-3}$, the oil bursts occur simultaneously, and the invading fluid interface is more compact over the scale of multiple pores. In both cases, we observe a ~1$\mu$m-thick layer of the wetting fluid coating the bead surfaces after oil invasion, indicated in the last frame of each sequence. Scale bars are 200$\mu$m.

**Figure 4. Pore-scale dynamics of secondary imbibition depend strongly on *Ca*.** Images show multiple frames, taken at different times after drainage, of a single optical slice. The slice is 11$\mu$m thick and is imaged within a porous medium comprised of beads with average radius $a = 75\mu$m, with cross-sectional width 3mm and height 3mm. The bright areas show the dyed wetting fluid; the dark circles are the beads, while the additional dark areas are the undyed oil being displaced from the pore volume. Direction of bulk wetting fluid flow is from left to right. The final frame shows the unchanging steady state. Labels show time elapsed after first frame. (a) At low $Ca = 7 \times 10^{-6}$, the wetting fluid pinches off the oil at multiple nonadjacent constrictions, then displaces the oil from the surrounding pores. The wetting fluid eventually flows through a tortuous, continuous network of filled pores, forming many trapped oil ganglia. (b) At high $Ca = 6 \times 10^{-4}$, the occurrence of oil pinch-off is reduced, and the wetting fluid displaces the oil from the pores, leaving a few small oil ganglia trapped within the medium. Scale bars are 200$\mu$m.

**Figure 5. 3D structure of trapped oil ganglia is strongly dependent on flow history.** 3D reconstructions of the steady-state pore-scale configurations of trapped oil ganglia (red) for porous media comprised of beads with average radius $a = 32\mu$m, with cross-sectional area 910$\mu$m x 910$\mu$m. Beads are not shown for clarity. Direction of bulk flow is from left to right. (a) At low $Ca = 8 \times 10^{-6}$, the trapped ganglia are ramified and can span many pores; (b) At high $Ca = 3 \times 10^{-4}$, the trapped ganglia are smaller and less ramified.

**Figure 6. Statistics of individual oil ganglion volumes, and total amount of trapped oil, are strongly dependent on flow history.** (a) Cumulative probability distribution functions of trapped ganglion volumes calculated from 3D reconstructions, with curves shifting to the left for increasing *Ca*; (b) Total residual oil saturation calculated from 3D reconstructions, showing that increasing amounts of oil are mobilized from the porous medium as *Ca* increases. Error bars indicate standard deviation of residual oil saturation measured at different locations of the same porous medium.

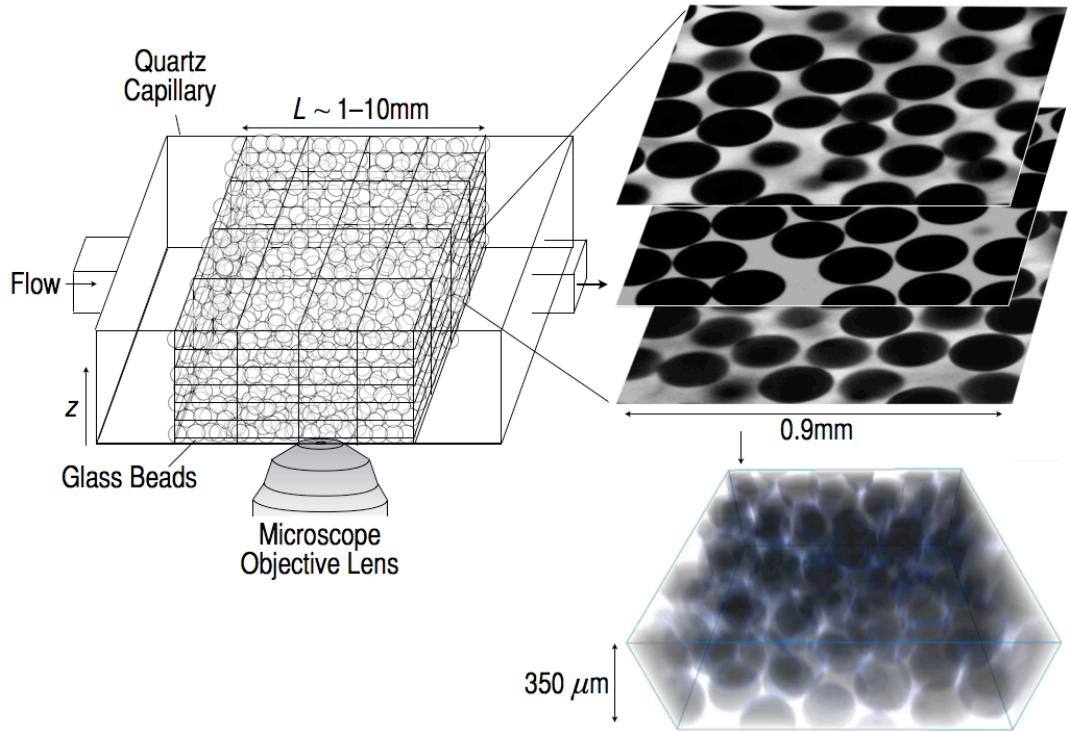

**Figure 1.**

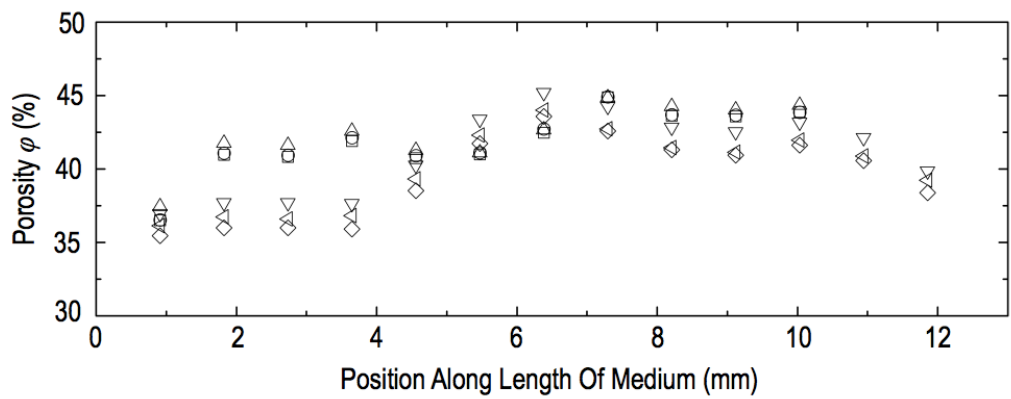

**Figure 2.**

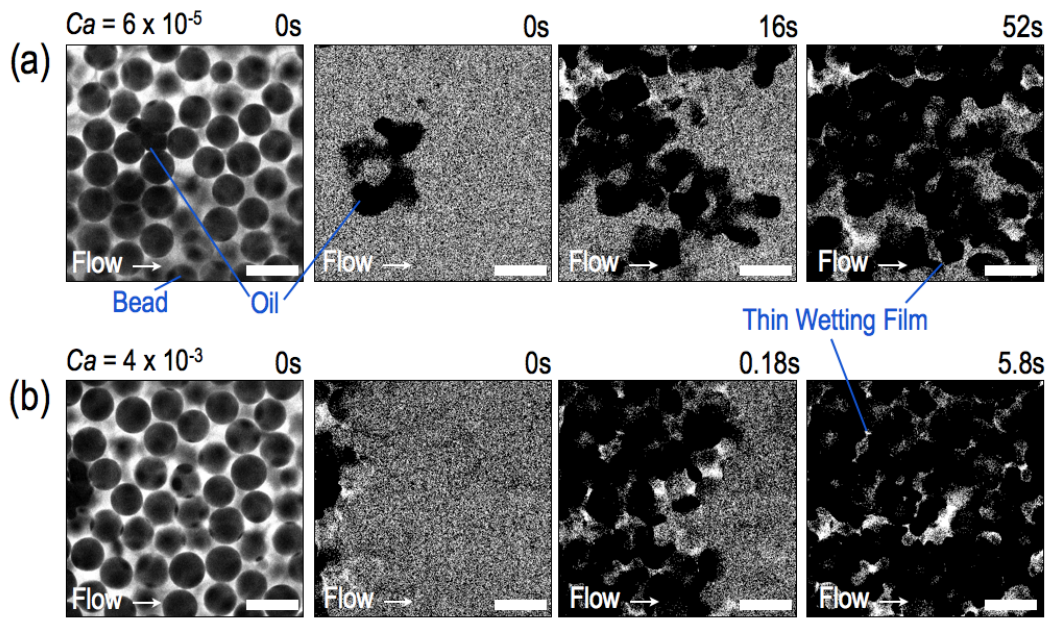

**Figure 3.**

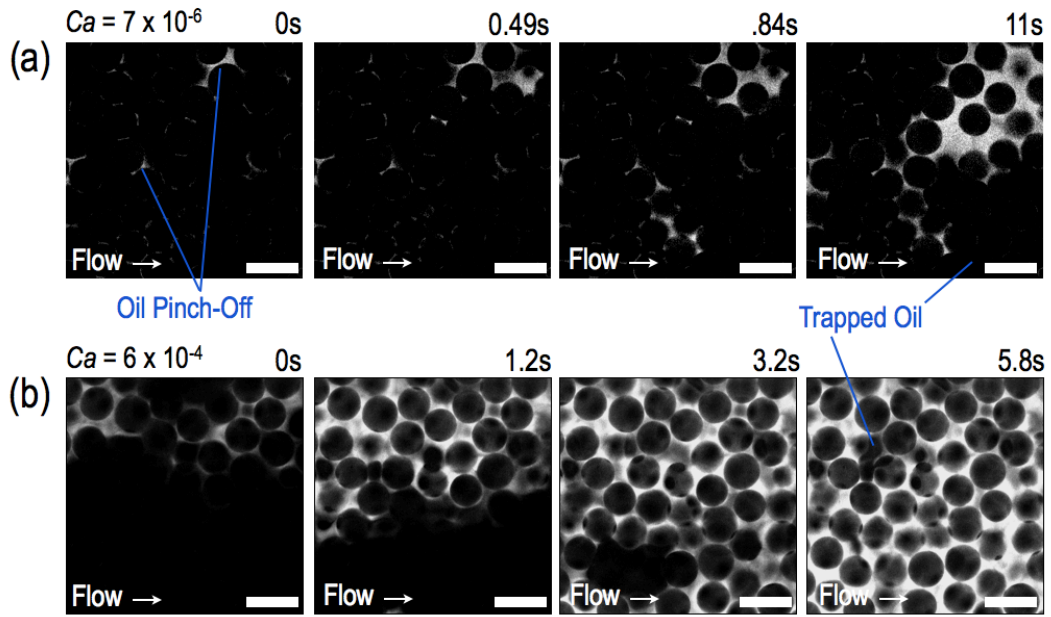

**Figure 4.**

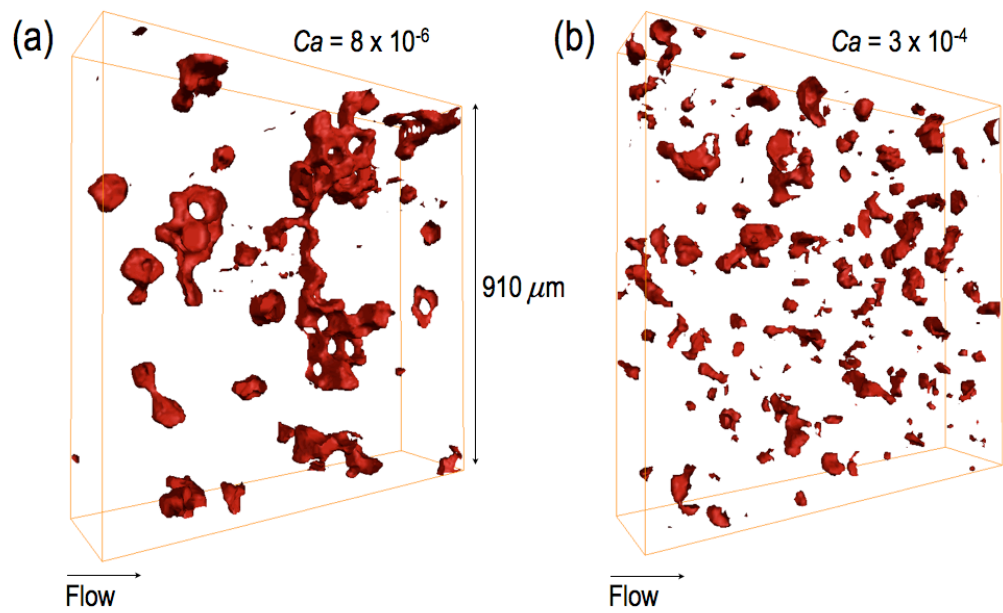

**Figure 5.**

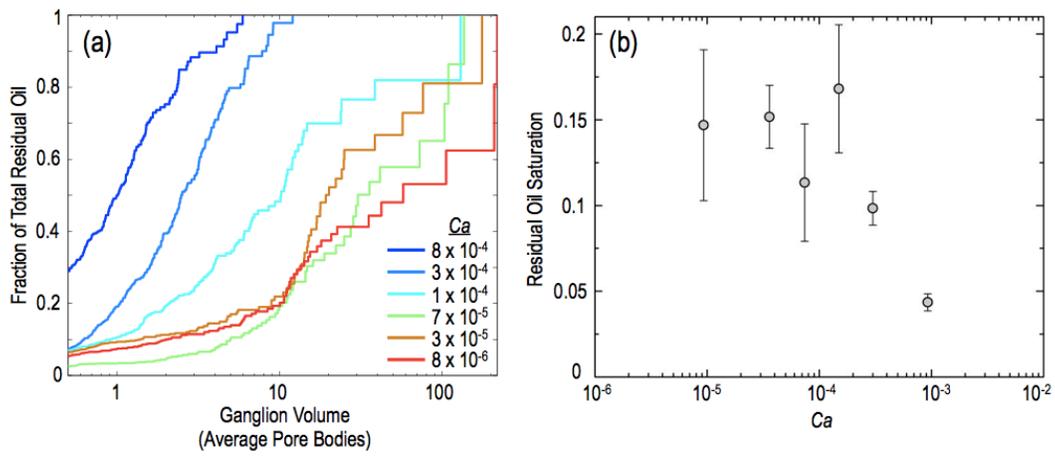

**Figure 6.**